\documentclass[%
 reprint,
 amsmath,amssymb,
 aps,prl
]{revtex4-1}

\usepackage{graphicx}
\usepackage{dcolumn}
\usepackage{bm}
\usepackage[utf8]{inputenc}
\usepackage[T1]{fontenc}
\usepackage{textcomp}
\usepackage{gensymb}

\usepackage{xr}
\begin{document}


\title{Strand plasticity governs fatigue in colloidal gels}

\author{Jan Maarten van Doorn}
 \altaffiliation{These authors contributed equally}
\author{Joanne E. Verweij}%
 \altaffiliation{These authors contributed equally}

\author{Joris Sprakel}
\author{Jasper van der Gucht}
\email{jasper.vandergucht@wur.nl}
\affiliation{%
 Physical Chemistry and Soft Matter, Wageningen University, Stippeneng 4, 6708 WE, Wageningen, The Netherlands
}%
%


\date{\today}

\begin{abstract}
Repeated loading of a solid leads to microstructural damage that ultimately results in catastrophic material failure. While posing a major threat to the stability of virtually all materials, the microscopic origins of fatigue, especially for soft solids, remain elusive. Here we explore fatigue in colloidal gels as prototypical inhomogeneous soft solids by combining experiments and computer simulations. Our results reveal how mechanical loading leads to irreversible strand stretching, which builds slack into the network that softens the solid at small strains and causes strain hardening at larger deformations. We thus find that microscopic plasticity governs fatigue at much larger scales. This gives rise to a new picture of fatigue in soft thermal solids and calls for new theoretical descriptions of soft gel mechanics in which local plasticity is taken into account.  
\end{abstract}

\maketitle

\begin{figure}[bt!]
\centering
\includegraphics[width=\linewidth]{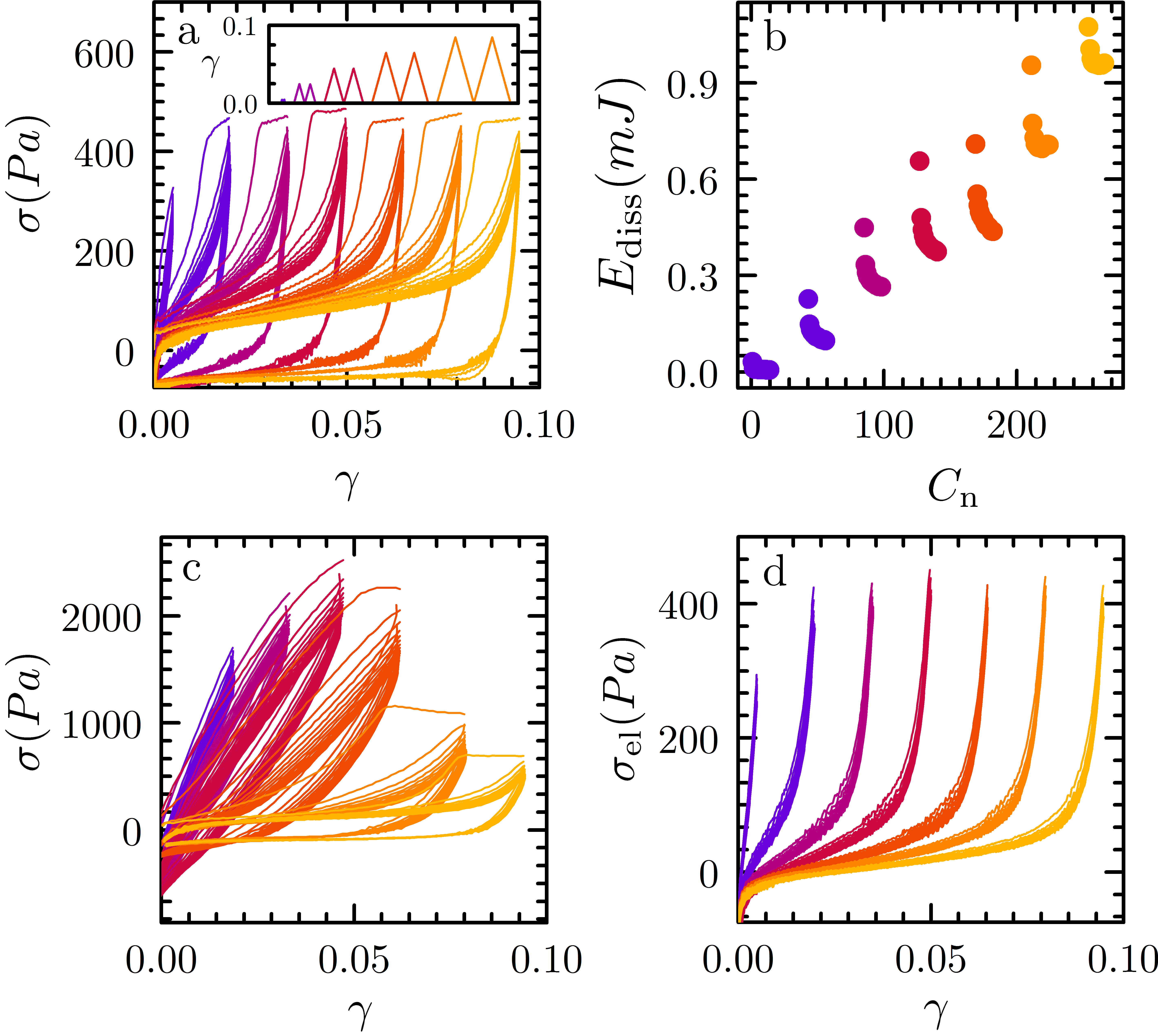}
\caption{(color online) a) Stress-strain curves as result of the strain profile depicted in the inset with $\dot{\gamma}$=10$^{-2}$ s$^{-1}$. From blue to yellow: $\gamma_{\text{max}}=$ 0.005,    0.02, 0.035, 0.05, 0.065, 0.08, 0.095 (note that only a selection of the sets is shown). b) Dissipated energy for every cycle for $\dot{\gamma}$=10$^{-2}$ s$^{-1}$. Colors indicate $\gamma_{\text{max}}$ which are the same as in (a)  c) Stress-strain curves for $\dot{\gamma}$=10$^{-1}$ s$^{-1}$, from blue to yellow: $\gamma_{\text{max}}=$ 0.005,    0.02, 0.035, 0.05, 0.065, 0.08, 0.095. d) Elastic midlines for curves in (a).}
\label{fig1}
\end{figure}

\begin{figure}[bt!]
\centering
\includegraphics[width=\linewidth]{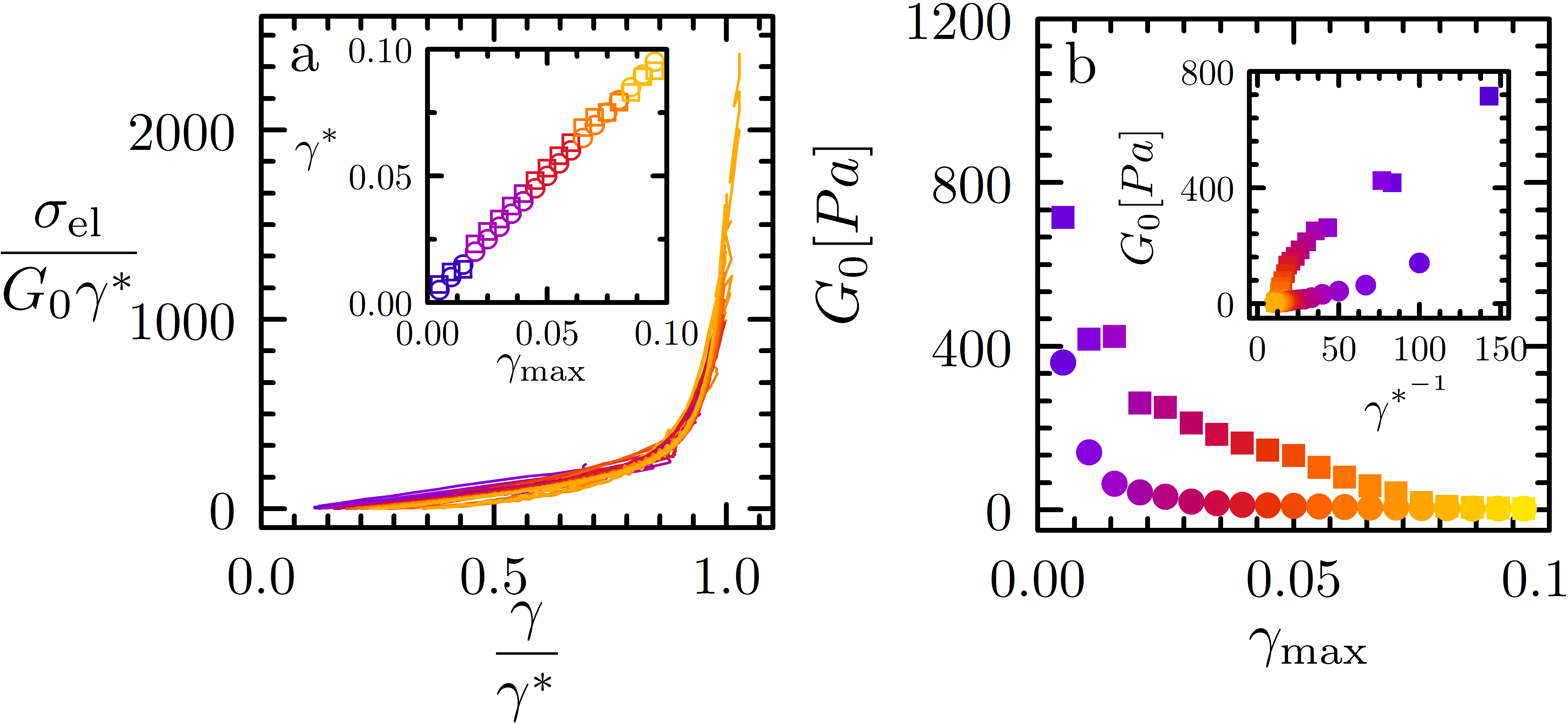}
\caption{(color online) a) Collapse of every second $\sigma_{\text{el}}$ elastic midline for both strains rates. The inset show the development of the $\gamma^{*}$ b) The initial modulus as function of $\gamma_{\text{max}}$ for $\dot{\gamma}$=10$^{-1}$ s$^{-1}$ (squares) and $\dot{\gamma}$=10$^{-2}$ s$^{-1}$ (circles). The inset depicts the dependence of $G_{0}$ on the inverse of $\gamma^{*}$ for $\dot{\gamma}$=10$^{-1}$ s$^{-1}$ (squares) and $\dot{\gamma}$=10$^{-2}$ s$^{-1}$ (circles).}
\label{fig2}
\end{figure}

The application of repeated load to a solid material can lead to the erosion of its microstruture, in a process that is known as fatigue. While the initial stages of this process often go unnoticed, the gradual accumulation of damage that results can ultimately lead to the sudden and catastrophic failure of the material.  Understanding the microscopic origins of fatigue is therefore crucial for the reliable prediction of a material's lifetime and for the development of strategies to improve mechanical stability. In  materials such as steel and concrete, fatigue is characterized by the accumulation and growth of small micro-cracks~\cite{Lee2004299,klesnil1992fatigue}. However, the mechanisms of fatigue in disordered soft materials are much less understood. A prototypical class of soft heterogeneous solids is comprised of colloidal gels.  These are non-equilibrium structures consisting of aggregated colloidal particles that form a sample-spanning network~\cite{trappe2001jamming}.\\
The arrested dynamics of the aggregated particles lead to solid-like behaviour, with elastic properties that are determined  by the structure and connectivity of the network~\cite{Krall1998, doi:10.1122/1.550812}. When subjected to a large enough stress, colloidal gels will eventually fluidize or fracture, often after a latent  period of apparent stability~\cite{Gopala2007,Bartlett2012, C5SM02587G,Kim2014, PhysRevLett.106.248303,Leocmach}. The microstructural changes responsible for this delayed failure have been attributed to the brittle-like rupture of network strands due to force-activated breaking of interparticle bonds~\cite{C2SM06723D,Colombo2014,Brice2017}. Such models assume that no restructuring of the network due to plastic particle rearrangements takes place. Yet, it is known that such rearrangements do occur~\cite{Landrum2016,Roy2016}, even for colloidal gels in rest~\cite{PhysRevLett.118.188001}, where they lead to aging, coarsening, and slow relaxation of internal stresses~\cite{PhysRevLett.84.2275, PhysRevE.80.010404,PhysRevLett.110.198301}. While it is known that under large deformation aggregates break into smaller clusters\cite{PhysRevE.80.051404}, the response of aggregated structures to repeated small deformations remains unclear. To establish a link between the stability of colloidal gels and their microstructure, it is therefore needed to understand the role of plasticity in gel failure and fatigue.\\
In this Letter, we report  fatigue measurements on model colloidal gels subjected to cyclic loading. By combining experiments and computer simulations, we show that the gradual weakening that occurs in these gels is due to plastic deformations within individual gel strands. Our results thus shed new light on the mechanism of damage accumulation in colloidal gels, and suggest that the current  models for colloidal gel failure must be revised to take this plastic softening into account.
 We investigate colloidal gels consisting of monodisperse polystyrene particles synthesized as described in ref. ~\cite{Appel2015}, with a volume fraction $\phi=$0.18. The particles have a radius $r=45$ nm as determined by dynamic light scattering. Attraction is induced by coating the particles with a thermo-responsive surfactant as synthesized in ref.~\cite{Kodger2013}, at a concentration of 1 g/L. To screen electrostatic repulsion between the particles, 100 mM NaCl is added to all samples. Rheological measurements are performed with a stress-controlled rheometer (MCR-501, Anton Paar) with a concentric cylinder geometry (CC10/Ti). The gels are formed in situ by heating the samples to 45$\degree$C, above the critical aggregation temperature of the surfactant, which results in gels with thick strands, each composed of many particles in its cross-section, in which it is established that significant rearrangements occur \cite{Kodger2013}. To minimize initial transient effects, samples are equilibrated for 1h before initiating measurements. Fatigue in the gels is studied by cyclically deforming the samples with a saw-tooth strain profile (Inset Fig.\ref{fig1}a). After 14 cycles at the same strain amplitude $\gamma_\text{max}$, we allow the sample to rest for 120 s, before starting a new set of deformation cycles at a higher strain amplitude, with increment $\Delta\gamma_\text{max}=0.005$.\\ 
For the smallest amplitude the stress-strain curve exhibits a linear elastic response (Fig. \ref{fig1}a). For larger strain amplitudes, however, the stress increases non-linearly with increasing strain and shows a hysteresis loop, which indicates dissipative losses during the deformation cycle ~\cite{doi:10.1122/1.4872059}. The first cycle for every strain amplitude differs qualitatively from the subsequent cycles: with increasing strain, the stress increases strongly, until a threshold value of approximately 470 Pa is reached, after which it levels off, indicating that the material undergoes plastic flow at this stress level due to local yielding. For every subsequent cycle, the observed stress is lower than that in the first cycle, signalling a progressive, irreversible weakening of the material. The dissipated energy is highest in the first cycle for a given strain amplitude and then gradually decreases to a plateau value as the stress-strain curve approaches a limit cycle (Fig. \ref{fig1}b). This limiting dissipated energy reflects the viscoelastic dissipation in the network due to solvent flow through the network or to reversible rearrangements, while the additional dissipation in the first cycle reflects the irreversible plastic deformation that occurs during loading of the gels.
Because the timescale at which plastic rearrangements take place can be relatively long for these strongly aggregated particles, we expect the amount of plasticity to depend on the loading rate. Indeed, when we increase  $\dot{\gamma}$ by a factor of 10, to $10^{-1}$ s$^{-1}$, we observe a much more elastic response, with the onset of non-linear plastic behaviour shifted to much larger strains and stresses (Fig. \ref{fig1}c). \\
While our data demonstrate the importance of plasticity for  fatigue in colloidal gels, the microscopic nature of this plastic deformation remains to be uncovered.  Given that colloidal gels are networks of connected strands consisting of aggregated particles,  the observed weakening must be caused either by the rupture of gel strands, leading to a decrease in network connectivity, or by softening of the gel strands, leading to a lower effective spring constant of the strands. To identify which of these scenarios is the dominant one, we analyze the mechanical response in more detail. First, we disentangle the elastic and viscous contributions to the mechanical response, by averaging the loading and unloading curve for each cycle~\cite{Munster2013}. This averages out the viscous contribution to the stress, so that only the elastic stress $\sigma_\text{el}$ remains (Fig. \ref{fig1}d). For the smallest strain amplitude, the stress strain response is linear; however, at higher strains, when the gels have undergone plastic deformation, the curves become strongly non-linear. The shapes of the non-linear response of all cycles are very similar. 
After an initial linear response, characterized by a linear modulus $G_0$, the gels show pronounced strain hardening:  at a characteristic strain amplitude $\gamma^*$, there is a sharp upturn of the stress after which the stress increases as a power law with strain, $\sigma\sim\gamma^x$, with an exponent $x\approx$ 2 that does not vary strongly between the different cycles. The linear modulus that characterizes the initial slope of the stress-strain curves decreases with increasing strain amplitude, signaling the progressive weakening of the gels resulting from the gradual erosion of the network structure during the fatigue cycles. We obtain $\gamma^{*}$ and $G_{0}$ by superimposing the different stress-strain curves for both strain rates by plotting the normalized stress, $\sigma_\text{el}/G_0\gamma^*$ as a function of the rescaled strain $\gamma/\gamma^*$ (Fig. \ref{fig2}a).\\
The excellent collapse indicates that the physical mechanism that underlies the mechanical response of the gels remains the  same during the fatigue cycles. We find a linear increase of $\gamma^*$ with increasing maximum strain amplitude  (inset Fig. \ref{fig2}a), indicating that the strain hardening response is delayed by the fatigue process. 
We attribute this linear increase of $\gamma^*$ with increasing $\gamma_\text{max}$ to the irreversible stretching of strands in the colloidal network during the regime of plastic flow, leading to the build-up of slack in the strands. During the next cycle the slack induced by the previous cycles is pulled out first, which results in little resistance and explains the initial soft linear elastic response of the gels. We find that the introduced slack leads to a softening of the gel, with $G_0\sim (\gamma^*)^{-1}$ for the low strain rate (inset Fig. \ref{fig2}b), suggesting a strand spring constant that is inversely proportional to extent by which it has become stretched. A similar scaling between spring constant and chain length is found in the entropic spring model for polymer chains \cite{rubinstein2003polymer}. For the higher strain rate, $G_0$ decreases more strongly at high strain amplitudes, which may indicate rupture of gel strands. When the strands are pulled taut, the resistance to further stretching increases strongly as the chain entropy vanishes and the physical bonds between the particles become perturbed. This results in the observed strain hardening in our colloidal gels. We note that these phenomena are reminscent of observations made for networks of biopolymer bundles ~\cite{Munster2013}.\\ 
\indent 
\indent To verify that lengthening of gel strands is indeed responsible for the plastic response of colloidal gels, we need detailed information at the individual strand level. Since this is extremely difficult to obtain experimentally for our system, we  perform Brownian Dynamics computer simulations on single gel strands. The strands are composed of 256 particles of diameter $a$ interacting through a Morse potential~\cite{PhysRevLett.114.258302}, with interaction range parameter $\rho_{0} = 33$ corresponding to a well width of approximately $\Delta = 0.3a$ and interaction strength $\beta\epsilon = 10$. Similarly to the experimental protocol, the gel strands are deformed cyclically with a saw-tooth strain profile (see supplemental material for further details) ~\footnote{See Supplemental Material at [URL will be inserted by publisher]}.\\ 
\indent To monitor the plastic rearrangements that occur in the gel strand during the strain cycles, we quantify the average plastic strain for each particle in an oscillation $c_n$ as
\begin{equation}
m_{i}(c_{n})= \frac{1}{\mathcal{N}_ia^2} \sum\limits_{j=1}^{\mathcal{N}_i} \langle\left({\bf r}_{ij}(0)-{\bf r}_{ij}(t)\right)^2 \rangle
\end{equation}
where ${\bf r}_{ij}(0)$ and ${\bf r}_{ij}(t)$ denote the separation vector between particle $i$ and neighbouring particles $j$ at the start of the cycle and after a time $t$, respectively, $\mathcal{N}_i$ is the number of nearest neighbours of particle $i$, and where the average is taken over the entire oscillation.\\
\indent During the strain cycles, the amount of plastic strain gradually increases, with most of the plastic rearrangements occurring during the first cycle. For larger strain amplitudes, the amount of plastic strain is also larger (SI Fig. 2a). Interestingly, the plastic rearrangements do not occur homogeneously in the gel strand, but are strongly localized to specific regions (Fig. 3). This leads to the formation of thicker and thinner regions in the gel strand, reminiscent of the Rayleigh-Plateau instability in liquid jets, which highlights the arrested liquid state of colloidal gels~\cite{lu2008gelation}. This is further corroborated by looking at the average number of bonds per particle, which gradually increases during the oscillations (SI Fig. 2c), suggesting that fatigue in colloidal gels is reminiscent of activated aging, in which the non-equilibrium gel structure tends to coarsen by increasing the number of bonds~\cite{Moghimi2017}.\\
\indent At high strains, the localization of plastic deformation ultimately leads to rupture of the gel strand at the weakest spot, i.e. at a local necking region (Fig. 3b). The number of ruptured strands increases with increasing strain amplitude and reaches about 65 \% for a strain of 0.06. Since our aim is to focus on the plastic rearrangements in the strands that precede strand rupture, we exclude the broken strands from further analysis. \\
\indent In addition we deform the gel strands at higher strain rates. Increasing the strain rate 100 times shows a considerably higher number of ruptured strands (SI Fig. 5). From the average plastic deformation per oscillation cycle (SI Fig. 6) as a function of increasing strain rate we observe a clear decline in the plasticity of the gel strands. This data suggest that colloidal gels with hardly any options to deform plastically will rupture in a brittle fashion. Macroscopically, the the extended linear regime in Figure \ref{fig1}c signals the onset of a transition to brittle failure. This is supported by the fact that $G_{0}$ deviates from the linear dependence on $(\gamma^{*})^{-1}$ at $\dot{\gamma}$=10$^{-1}$ s$^{-1}$ , suggesting a larger part of the damage is caused by brittle fracture (inset Fig \ref{fig2}b).\\ 
\begin{figure}[bt!]
\centering
\includegraphics[width=\linewidth]{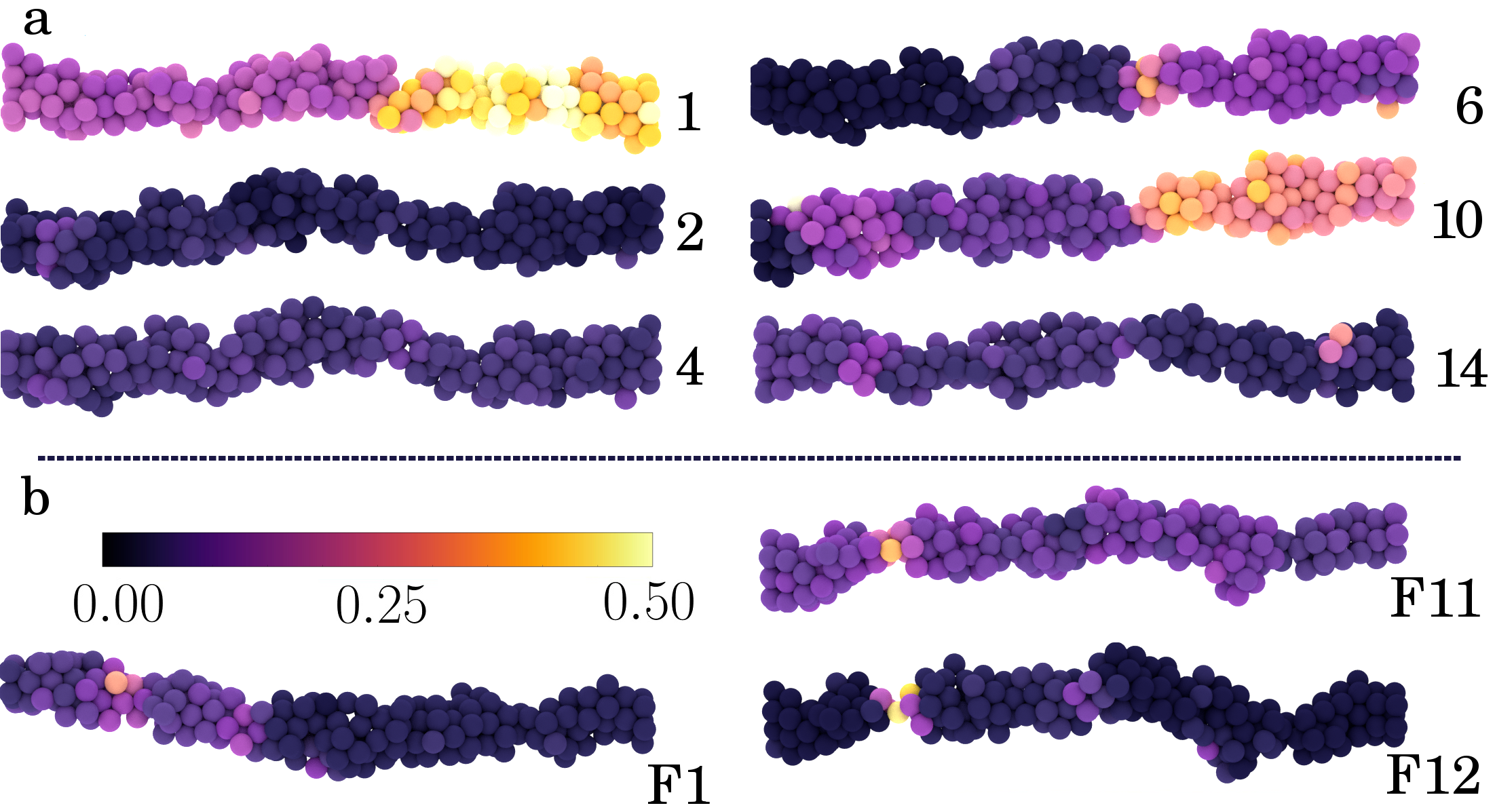}
\caption{(color online) (a) Visual representation of the non-cumulative average plastic deformation per particle in oscillation cycles 1,2,4,6,10 and 14 of a single gel strand ($\gamma_{\text{max}} = 0.04$, data SI Fig 3.). The color bar indicates the non-cumulative plastic deformation $m_{i}(c_{n})$ per particle in each cycle from low (purple) to high (yellow). The cumulative plastic deformation of this gel strand is shown in SI Fig. 7. (b) Plastic deformation in a gel strand before fracture (F11) and and after fracture (F12).}
\label{fig3}
\end{figure}   
\indent To connect our simulation results to the experimental findings, we calculate the force needed to deform the strand as a function of the strain~\footnote{The force $f$ on the simulation walls is related to the stress measured in the experiment as $\sigma \approx \frac{f}{\xi^2}$, with $\xi$ the mesh size of the network.}. The force-strain curves for a single gel strand show features that are very similar to the macroscopic curves measured experimentally (Fig. \ref{fig4}a). Similarly to the experimental curves, we find that the first force-distance curve for each strain amplitude differs qualitatively from the subsequent cycles, showing a plateau above a critical force that indicates plastic deformation. This plastic deformation reflects the particle rearrangements shown in Fig. \ref{fig3}. \\
\indent As a consequence, the dissipated energy  is highest in the first oscillation cycles and reaches a plateau after a few oscillations (Fig. \ref{fig4}b, SI Fig. 3b). The dissipated energy can be correlated directly to the number of inter-particle bonds that are broken in the gel strands, which follows the same trend (SI Fig. 2b).  As the gel strands are several particle diameters wide (Fig. \ref{fig3}), the breaking of a single bond does not immediately lead to rupture of the whole strand \cite{C2SM06723D}. The overall integrity of the strand is maintained by adjacent bonds and due to thermal fluctuations, new bonds can form ~\cite{Munster2013, PhysRevLett.110.198301, PhysRevLett.106.248303}. This provides a mechanism for plastic deformation of the gel strands.\\  
\begin{figure}[bt!]
\centering
\includegraphics[width=\linewidth]{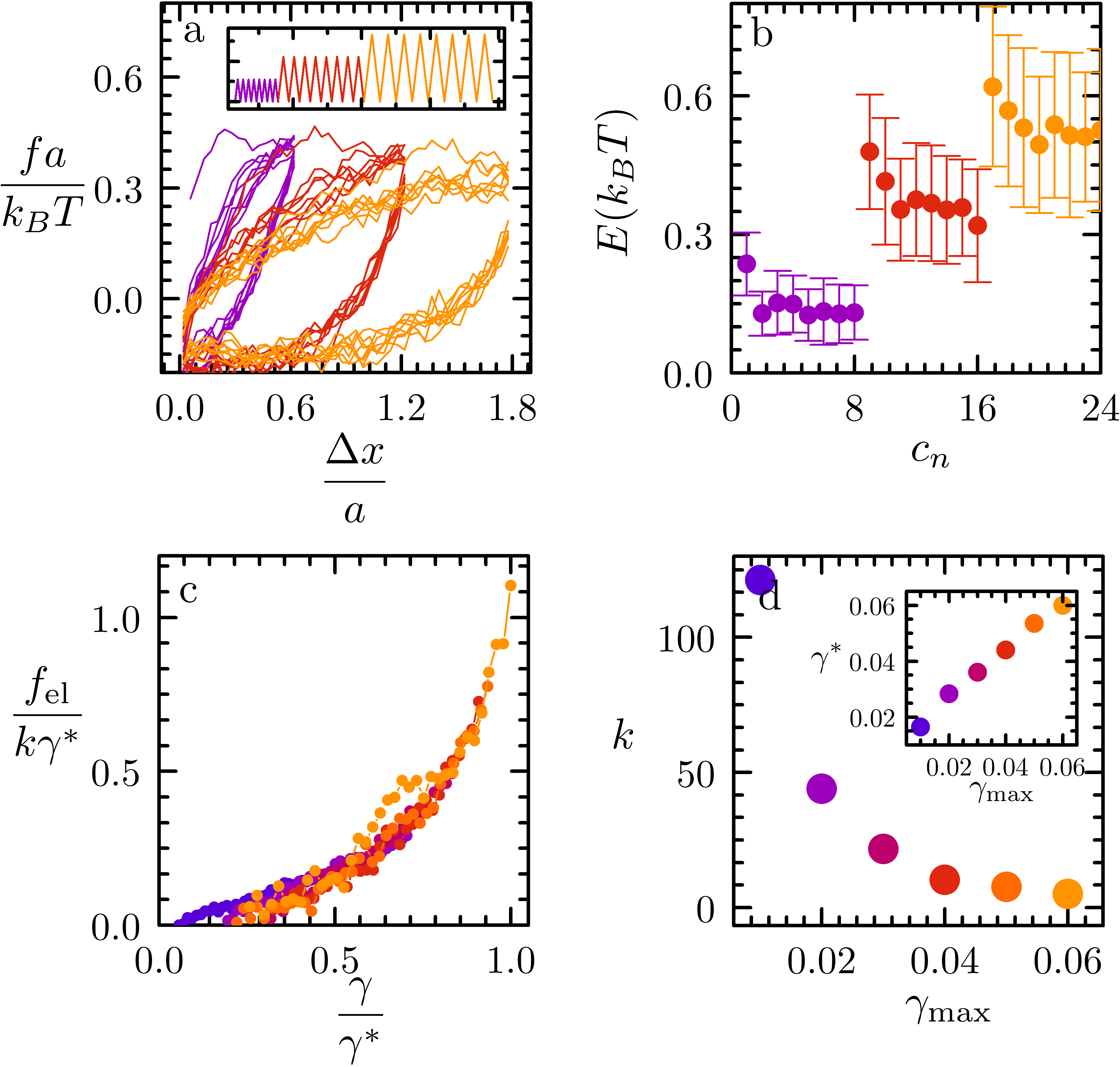}
\caption{(color online) (a) Force-distance curves for BD simulations of 24 (8x3) oscillations of a single gel strand at strain amplitudes (purple to red) $\gamma=$ 0.02, 0.04 and 0.06. (b) Dissipated energy per oscillation for strain amplitudes $\gamma=$ 0.02, 0.04 and 0.06, obtained by integration of the force-distance curves. (c) Collapse of the average force-distance curves (positive parts of the loading and un-loading curve) of the 4$^{th}$ oscillation cycle scaled by $\gamma^{*}$ on the x-axis and $k \cdot \gamma^{*}$ on the y-axis. Data is obtained from SI Fig. 3. (d) Spring constant $k$ (in units $k_BTa^{-1}$) as a function of $\gamma_{max}$; inset: $\gamma^{*}$ as a function of $\gamma_{max}$.}
\label{fig4}
\end{figure}
\indent To check whether the plastic particle rearrangements found in the simulations can provide an explanation for the experimentally observed fatigue response, we use a similar procedure to re-scale the force distance curves as in Fig. \ref{fig2}a. Again, we average the force in the loading and unloading curve and plot the rescaled elastic force $f_\text{el}/k \cdot \gamma^{*}$, with $k$  the fitted spring constant of the strand in the linear regime as a function of the rescaled strain $\gamma/\gamma^{*}$. This yields a curve that is very similar to the experimental one (Fig. \ref{fig4}c), with a linear response regime, followed by a strain-hardening response at higher strains. The onset of strain hardening shifts to higher strains with increasing strain amplitude (inset Fig. \ref{fig4}c), in agreement with experiments (inset Fig. \ref{fig2}a) and confirming our hypothesis of a gradual build-up of slack in the strands. Also the linear spring constant of the gel strands $k$ decreases  in a similar fashion with the strain amplitude $\gamma$ as the elastic modulus in the experiments (Fig. \ref{fig4}d).\\ 
\indent Since all broken strands are excluded from the analysis, the observed weakening in the simulated force-distance curves can be attributed completely to plastic particle rearrangements within the  strands. This suggests that also the    weakening observed at the macroscopic scale in our fatigue experiments can be explained by plastic deformation and local necking in individual strands, without the need to invoke rupture of strands.\\
\indent Our results highlight how fatigue in colloidal gels results from plasticity at much smaller scales. This feature results from the strongly hierarchical and multiscale structure of networks of colloidal particles. To date, strand plasticity has been overlooked in describing the mechanics of colloidal gels but has also received little attention as a possible mechanism of fatigue in a wider variety of heterogeneous solids, while our data clearly indicate its pivotal role in deciding the materials fate under repeated loading. Strand stretching and the build-up of slack has also been identified as a mechanism for strain softening in networks of biological fibers~\cite{Munster2013}. This raises the question whether localized plasticity, which remains obscured in macroscopic mechanical testing, may have a more universal role in the fatigue mechanisms of a wider class of inhomogeneous soft solids with a hierarchical microstructure. If so, a universal description of these effects could have pronounced implications for the predictability of the non-linear mechanical response of soft materials, which remains an open challenge in the field. Finally, while our work has focused on fatigue induced by external loading, other failure mechanisms driven by internal stresses are known to exist for these inhomogeneous thermal solids, such as ageing and syneresis~\cite{PhysRevLett.84.2275, PhysRevE.80.010404,PhysRevLett.110.198301}. The plasticity we describe here results from the rearrangement of particles by thermally-activated debonding~\cite{PhysRevLett.118.188001}, in which the mechanical stress imposes a directional bias that leads to irreversible strand stretching. Based on our observations here, we hypothesize that internal stress can give rise to similar effects, where e.g. contractile internal stresses could bias rearrangements that lead to isotropic condensation of the structure, ultimately resulting in syneresis. While this remains unexplored to-date, it could open the way to a universal description of the failure of these non-equilibrium solids. 

\subsection{Acknowledgement}
This work is part of the Industrial Partnership Programme Hybrid Soft Materials that is carried out under an agreement between Unilever Research and Development B.V. and the Netherlands Organisation for Scientific Research (NWO). JvdG acknowledges the European Research Council for financial support (ERC Consolidator grant Softbreak). The work of J.S. is part of the VIDI research programme with project number 723.016.001, which is financed by the Netherlands Organisation for Scientific Research (NWO).

\section*{References}
\bibliography{PSRheoArticle}

\end{document}